\documentclass[aps,prb,reprint,superscriptaddress,longbibliography]{revtex4-2}
\usepackage{graphicx}
\usepackage{amsmath}
\usepackage{amssymb}
\usepackage{bm}
\usepackage[utf8]{inputenc}

\usepackage{color}

\begin{document}

\title{Optically induced spin Hall current in Janus transition-metal dichalcogenides}

\author{Tomoaki Kameda}
\affiliation{Department of Nanotechnology for Sustainable Energy,
  School of Science and Technology, Kwansei Gakuin University,
  Gakuen-Uegahara 1, Sanda 669-1330, Japan} 
\author{Katsunori Wakabayashi}
\affiliation{Department of Nanotechnology for Sustainable Energy,
  School of Science and Technology, Kwansei Gakuin University,
  Gakuen-Uegahara 1, Sanda 669-1330, Japan} 
\affiliation{Center for Spintronics Research Network (CSRN), Osaka University, Toyonaka 560-8531, Japan}
\affiliation{National Institute for Materials Science (NIMS), Namiki
  1-1, Tsukuba 305-0044, Japan}

\begin{abstract}
Monolayer Janus transition-metal dichalcogenides (TMDCs), such as
WSeTe, exhibit Rashba-type spin-orbit coupling (SOC) due to broken out-of-plane
mirror symmetry. Here, we theoretically demonstrate that 
pure spin Hall currents can be generated under light
irradiation based on a tight-binding model.
Rashba-type SOC plays a crucial role in determining the spin polarization direction 
and enhancing the generation efficiency of pure spin Hall currents.
Our findings establish Janus TMDCs as promising materials 
for next-generation optospintronic devices. 
\end{abstract}

\maketitle

\section{Introduction}
Transition-metal dichalcogenides (TMDCs) 
provide a new platform for two-dimensional (2D) electronic
systems~\cite{Novoselov2005,Manzeli2017,Mak2014,Lv2015,Radisavljevic2011}. 
TMDCs offer various degrees of freedom, such as charge, valley, and
spin, which are essential for spintronic and optoelectronic
devices~\cite{Mak2010,Mak2016,Wang2012,Splendiani2010,Wolf2001,Long2019,Tongay2012,Gutierrez2013,Zhao2013}.
In particular, monolayer group-VI TMDCs (e.g., MoS$_2$,
WS$_2$, WSe$_2$) are semiconductors with a direct energy band gap.
These materials possess a hexagonal lattice structure with broken
in-plane inversion symmetry while retaining out-of-plane mirror
symmetry. 
Owing to their lattice structure and the presence of transition metal
atoms, 
Ising-type spin-orbit coupling (SOC) dominates in 
certain monolayer group-VI TMDCs, acting as an
effective Zeeman field that aligns the spins in the out-of-plane
direction~\cite{Zhu2011,Lu2015,Zhou2016,He2018}. This SOC plays a
crucial role in unique spin transport phenomena, such as the spin Hall
effect (SHE), facilitating the efficient generation and conversion of
spin
currents~\cite{Kormanyos2014,Qian2014,Feng2012,Safeer2019,Ominato2020}.

Recently, monolayer Janus TMDCs with a chemical formula $MXY$,
e.g. $M$: Mo, W; $X, Y$: S, Se, Te, have been successfully 
synthesized~\cite{Lu2017,Zhang2017,Zhang2022}.  
A key feature of monolayer Janus TMDCs is the lack of both in-plane
inversion symmetry and out-of-plane mirror symmetry owing to the 
substitution of one chalcogen layer with another different chalcogen atom.
In Janus TMDCs, the difference in electronegativity
between the two different chalcogen atoms generates
an internal electric field perpendicular to the
plane. As a result, monolayer Janus TMDCs exhibit unique properties,
including piezoelectricity~\cite{Dong2017,Guo2017}, long exciton
lifetime~\cite{Jin2018,Zheng2021}. 
Furthermore, Janus TMDCs possess Rashba-type SOC as well as Ising-type SOC~\cite{Yao2017,Hu2018,Chen2020_2}. 
Unlike Ising-type SOC,
Rashba-type SOC aligns spin polarization parallel to the in-plane direction~\cite{Freimuth2021}.
This enhances SHE, facilitating 
efficient spin Hall current generation and 
conversion, which are key factors for spintronic applications. 

In this paper, we theoretically demonstrate the generation of pure
spin Hall currents in monolayer Janus TMDCs under light irradiation.  
We numerically calculate the spin-dependent optical conductivity of
monolayer Janus TMDCs using the Kubo formula based on an effective
tight-binding model (TBM) with the inclusion of Rashba-type SOC effects.  
For monolayer Janus TMDCs, Rashba-type SOC plays a crucial role in
generating pure spin Hall currents by determining the direction of spin
polarization and enhancing their generation efficiency. 
Our results contribute to the development of optspintronics devices
utilizing 2D materials, enabling the control of spin polarization
direction via Rashba-type SOC arising from the crystal asymmetry.

This paper is organized as follows. 
In Sec.~\ref{sec_Electric properties}, we analyze the electronic
properties of monolayer Janus TMDCs using TBM with the inclusion of
Rashba-type SOC effects. The energy band structure exhibits spin
splitting, where spin polarizations oriente along the $x$ and $y$
directions. 
In Sec.~\ref{sec_spin-dependent-cond}, we calculate the spin-dependent
conductivity using the Kubo formula and demonstrate that pure spin Hall current
can emerge in monolayer Janus TMDCs under both direct current (DC)
limit conditions and light irradiation. We discuss
the in-plane SHE in monolayer Janus TMDCs. 
Sec.~\ref{sec_conclusion} provides the summary. 
In Appendix~A, we provide detailed expressions for the matrix
elements of the effective Hamiltonian.
In Appendix~B, we analyze the
symmetry of the integrands of conductivities in the Brillouin zone (BZ). 
In Appendix~C, using Neumann's principle, we present the properties of
symmetry in the spin Hall conductivity.
In Appendix~D, we derive the optically generated spin current by
circularly polarized light irradiation.
In Appendix~E, we discuss the temperature
dependence of the spin-dependent optical Hall conductivity. 

\section{Effective Tight-Binding Model}\label{sec_Electric properties}
We demonstrate that pure spin Hall currents can be generated in
monolayer Janus TMDCs under light irradiation. 
Specifically, we use WSeTe as an
example of a Janus TMDC in this study because it exhibits strong Rashba-type
SOC. In this section, we describe the structure of monolayer Janus
TMDCs and analyze their electronic states using an effective TBM
that incorporates SOC effects. 
\subsection{Lattice structure}
\begin{figure*}
  \centering
  \includegraphics[width=170mm]{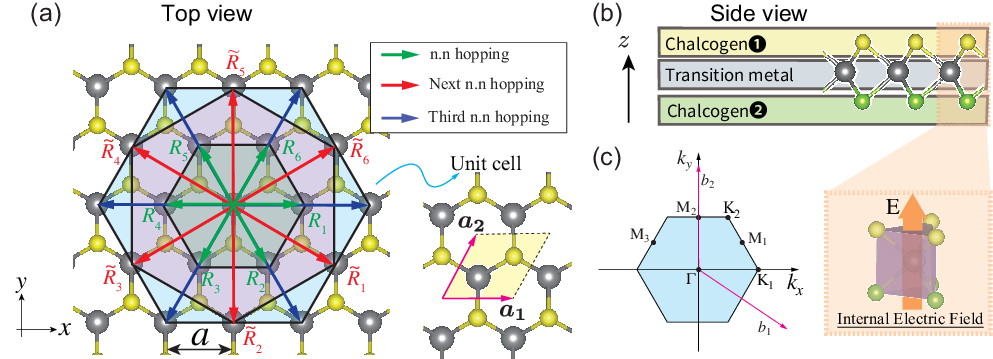}
  \caption{
    (a) Top view of monolayer Janus TMDCs, illustrating the
    nearest-neighbor, next-nearest-neighbor and third-nearest-neighbor
    hopping vectors, (green, blue, and red arrows, respectively).
    The yellow-shaded rhombus indicates the unit cell of the Janus
    TMDCs lattice structure. $\bm{a}_1 = (a, 0)$, $\bm{a}_2 =
    (\frac{a}{2},\frac{\sqrt{3}a}{2})$ are primitive vectors. 
    (b) Side view of monolayer Janus TMDCs, showing the transition
    metal layer sandwiched between two different chalcogen layers,
    labeled as Chalcogen 1 and Chalcogen 2.
    Due to the presence of different chalcogen atoms, an internal electric field
    $E$ is induced perpendicular to the plane. (c) First
    Brillouin zone (BZ) of monolayer Janus TMDCs with high-symmetry
    points $\Gamma$, $K_1$, $K_2$, $M_1$, $M_2$ and $M_3$.
  }\label{janus_tmdc_structure}
\end{figure*}

Figures~\ref{janus_tmdc_structure}(a) and (b) show the top and side views
of monolayer Janus TMDCs, respectively. Monolayer Janus TMDCs belong to
the space group $C_{3v}$, which consists of a transition metal layer
sandwiched between two layers of different chalcogen atoms. Therefore,
they have a hexagonal lattice structure
but lack out-of-plane mirror symmetry, distinguishing them from
conventional TMDCs [Fig.~\ref{janus_tmdc_structure}(b)]. This unique structure
generates an internal electric field perpendicular to the
materials. Additionally, Fig.~\ref{janus_tmdc_structure}(a) shows the unit
cell (the yellow shaded rhombus). Here, the magenta arrows are
primitive vectors, given as $\bm{a}_1 = (a, 0)$ and $\bm{a}_2 =
(\frac{a}{2},\frac{\sqrt{3}a}{2})$, where $a$ is the lattice constant:
$a=3.325$ \AA.  
Since the corresponding reciprocal lattice vectors are given as
$\bm{b}_1 = \frac{2\pi}{a}(1,-\frac{1}{\sqrt{3}}) $ and $\bm{b}_2 =
\frac{2\pi}{a}(0,\frac{2}{\sqrt{3}})$, the first BZ
of monolayer TMDCs form a hexagonal shape as shown in
Fig.~\ref{janus_tmdc_structure}(c). 

\subsection{Hamiltonian}
For WSeTe, we employ a multiorbital TBM that includes $d_{z^2}$,
$d_{x^2-y^2}$, and $d_{xy}$ orbitals of the W atom to describe the
low-energy electronic states of monolayer Janus TMDCs~\cite{Liu2013,Zhou2019,Zhou2020}.
The eigenvalue equation of the effective TBM is
\begin{equation}
\hat{H}(\bm{k})|u_{n\bm{k}}\rangle = E_{n}(\bm{k})|u_{n\bm{k}}\rangle,
\end{equation}
where $\bm{k} = (k_x, k_y)$ is the wave-number vector, $E_{n}(\bm{k})$
is the eigenvalue and $n = 1, 2, \ldots , 6$ denotes the band index.  
The eigenvector is defined as 
\begin{widetext}
\begin{equation}
|u_{n\bm{k}}\rangle = {(c_{n\bm{k},d_{z^2},\uparrow},
  c_{n\bm{k},d_{xy},\uparrow}, c_{n\bm{k},d_{x^2-y^2},\uparrow},
  c_{n\bm{k},d_{z^2},\downarrow}, c_{n\bm{k},d_{xy},\downarrow},
  c_{n\bm{k},d_{x^2-y^2},\downarrow})}^T, 
\end{equation}
where ${( \cdots )}^T$ indicates the transpose of the vector and
$c_{n\bm{k}\tau\sigma}$ means the amplitude at atomic orbital $\tau$
with spin $\sigma$ for the $n$th energy band at $\bm{k}$.  
The Hamiltonian with the Ising-type and Rashba-type SOCs can be written as
\begin{equation}
  \hat{H}(\bm{k}) = \hat{\sigma}_0 \otimes \hat{H}_{\text{TNN}}(\bm{k}) + \hat{\sigma}_z \otimes \frac{1}{2}\lambda_{\text{SOC}}\hat{L}_z - \hat{\sigma}_0 \otimes \mu \hat{I}_{3 \times 3} + \hat{H}_R(\bm{k}), 
  \label{eq:hamiltonian}
\end{equation}
\end{widetext}
with
\begin{equation}
\hat{H}_{\text{TNN}}(\bm{k}) = \begin{pmatrix}
V_0 & V_1 & V_2 \\
V_1^* & V_{11} & V_{12} \\
V_2^* & V_{12}^* & V_{22}
\end{pmatrix},
\label{eq:H_TNN}
\end{equation}
\begin{equation}
\hat{L}_z = \begin{pmatrix}
0 & 0 & 0 \\
0 & 0 & -2i \\
0 & 2i & 0
\end{pmatrix},
\end{equation}
and
\begin{equation}
\hat{H}_R(\bm{k}) = (f_x(\bm{k})\hat{\sigma}_y - f_y(\bm{k})\hat{\sigma}_x) \otimes \begin{pmatrix}
2\alpha_0 & 0 & 0 \\
0 & 0 & 0 \\
0 & 0 & 0
\end{pmatrix}.
\end{equation}
Here, $\sigma_0$ and $\hat{I}_{3\times 3}$ represent a $2\times 2$ and
$3\times 3$ identity matrix, respectively. $\hat{\sigma}_i$ is
$i$-component of Pauli matrices ($i=x, y, z$) and
$\lambda_{\text{SOC}}$ is the Ising-type SOC parameter.  
For monolayer WSeTe, we use $\lambda_{\text{SOC}} = 0.228\ \text{eV}$.
Note that we adopt the Ising-type SOC parameter from WSe$_2$ as reported in previous work~\cite{Liu2013}. 
Although WSeTe is a Janus material with broken out-of-plane symmetry,
its underlying electronic structure associated with Ising-type SOC
remains largely similar to that of WSe$_2$. Previous studies have shown
that Ising-type SOC is robust and only slightly affected by the
formation of a Janus structure~\cite{Zhou2019}.

The first term in Eq.~(\ref{eq:hamiltonian}) represents the
spin-independent component.
$\hat{H}_{\text{TNN}}(\bm{k})$ includes the electron hoppings only
among three $d$ orbitals of transition metal atoms, which are assumed
up to third-nearest neighbor (NN) sites.
As shown in Fig.~\ref{janus_tmdc_structure}(a),
green, red, and blue arrows indicate hopping vectors $\mathbf{R}_i (i = 1, 2, \cdots , 6)$ pointing to NN sites, 
the vectors $\tilde{\mathbf{R}}_j (j = 1, 2, \cdots , 6)$ pointing to next-NN sites, and the vectors $2\mathbf{R}_i$ pointing to third-NN sites, respectively.
The details of matrix elements $V_0$, $V_1$, $V_2$, $V_{11}$,
$V_{12}$, and $V_{22}$ can be found in
Appendix~\ref{sec_appendix_TBM}. 
The second term of Eq.~(\ref{eq:hamiltonian}) corresponds to the
Ising-type SOC, which arises from atomic spin-orbit interaction. 
In the third term of Eq.~(\ref{eq:hamiltonian}), $\mu$ denotes the chemical potential.
The fourth term of Eq.~(\ref{eq:hamiltonian}), $H_R(\bm{k})$, represents
the Rashba-type SOC, which arises due to the breaking of out-of-plane
mirror symmetry. $\alpha_0$ denotes the parameter that represents the
effective Rashba-type SOC strength for \{$d_{z^2}$\}. 
In this paper, we set $\alpha_0$ to $0.045\ \text{eV}$ for WSeTe,
a value that has been validated in previous studies on Janus TMDCs~\cite{Zhou2019}.
Rashba-type SOC for the \{${d_{xy}, d_{x^2 - y^2}}$\} is neglected because the spin splitting induced by Rashba-type SOC occurs mainly in the valence band around the $\Gamma$ point, which is dominated by the $d_{z^2}$ orbitals~\cite{Liu2013,Zhou2019,Yao2017}. 
If we define $(\alpha, \beta)=(\frac{1}{2}k_xa,\frac{\sqrt{3}}{2}k_ya)$, the functions $f_x(\bm{k})$ and $f_y(\bm{k})$ are given by
\begin{align}
f_x(\bm{k})& = \sin(2\alpha) + \sin(\alpha)\cos(\beta), \\
f_y(\bm{k})&= \sqrt{3}\sin(\beta)\cos(\alpha).
\end{align}

\begin{figure}
  \centering
  \includegraphics[width=85mm]{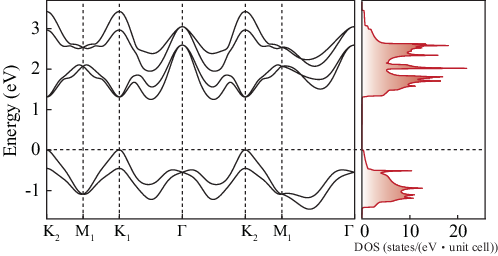}
  \caption{
    Energy band structure and DOS of monolayer WSeTe along
    high-symmetry points in 1st BZ.
  }\label{janus_tmdc_band}
\end{figure}
Figure~\ref{janus_tmdc_band} shows the energy band structure of
monolayer WSeTe along the line passing through the high-symmetry
points of the first Brillouin zone (BZ) and the corresponding density
of states (DOS). Here, valence-band maximum at $K$-point is set to
zero. Monolayer WSeTe is a semiconductor with a direct band gap of
$1.2$ eV at the $K_1$ and $K_2$ points. 

\subsection{Spin-polarized Fermi lines of hole-doped WSeTe}
\begin{figure*}
  \centering
  \includegraphics[width=170mm]{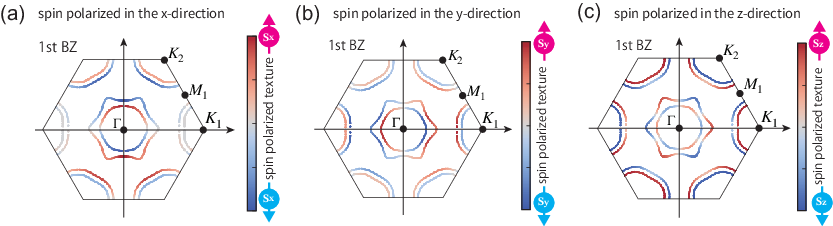}
  \caption{
Contour plot of Fermi lines with spin polarization of hole-doped
monolayer WSeTe. The Fermi energy is set $-0.85$ eV. 
Color indicates the strength of spin polarization 
along the (a) $x$, (b) $y$, and (c) $z$-direction.
}\label{janus_tmdc_spin_band}
\end{figure*}
Figures~\ref{janus_tmdc_spin_band}(a)-(c) represent spin-polarized
contour plots of the Fermi lines for hole-doped Janus TMDC, where
the Fermi energy is set to $-0.85$ eV.
Here, the color indicates the spin polarization in (a) $x$, (b) $y$,
and (c) $z$-directions, respectively. 
Figures~\ref{janus_tmdc_spin_band}(a) and (b) illustrate the
anisotropic spin polarization induced
by Rashba-type SOC due to the broken out-of-plane mirror symmetry.
Figure~\ref{janus_tmdc_spin_band}(c) shows the spin polarization
along the out-of-plane $z$-direction, driven by the broken inversion symmetry and strong Ising-type SOC. 
Conventional monolayer TMDCs exhibit spin polarization solely in the
out-of-plane direction~\cite{Habara2021}. In contrast, monolayer Janus
TMDCs exhibit both in-plane and out-of-plane spin polarization due to
the presence of both Rashba-type and Ising-type SOCs. 
The coexistence of in-plane and out-of-plane spin polarization in
monolayer Janus TMDCs enables versatile control of spin currents,
suggesting the potential for more advanced spintronic functionalities.

\section{spin-dependent conductivity}\label{sec_spin-dependent-cond}
In this section, we consider the spin-dependent optical conductivity
of WSeTe under linearly polarized light irradiation.
Here, the spin-dependent optical conductivity is numerically
calculated 
using Kubo formula based on an effective
TBM.~\cite{Guo2005,Sengupta2016,Vargiamidis2014,Tanaka2008,Ferreira2011,Morimoto2009,Yao2004,Li2012,Akita2020,Qiao2018} 
Under the application of an external electric field
$\bm{\mathcal{E}}(\omega)=\sum_j\mathcal{E}_j(\omega)\bm{e}_j$ ($j=x, y$) to the system, the
linear response of the 
spin current $\bm{J}^{\text{spin}{(k)}}(\omega)$ can 
be expressed as
\begin{equation}
  \bm{J}^{\text{spin}(k)}(\omega) = \bm{\sigma}^{\text{spin}(k)}(\omega)\bm{\mathcal{E}}(\omega),
\end{equation}
where the $\bm{\sigma}^{\text{spin}(k)}(\omega)$ is the spin-dependent
optical conductivity tensor for optical angular frequency $\omega$. The
superscript $\text{spin}(k)$ 
indicates the direction of spin
polarization. Thus, the spin current flowing along $i$ direction with the spin polarization $k$
can be expressed as  
\begin{equation}
  J^{\text{spin}(k)}_{i}(\omega) = \sum_{j}\sigma^{\text{spin}(k)}_{ij}(\omega)\mathcal{E}_j(\omega),
  \label{eq:spin-dependent-opt-current}
\end{equation}
with 
\begin{align}
  \sigma_{ij}^{\text{spin}(k)}(\omega)=\frac{i\hbar e}{{(2\pi)}^2}\int_{\text{BZ}}d^2\bm{k}\sum_{n\neq m}f_{nm}\nonumber \\
  \times \frac{\langle
    u_n(\bm{k})|\Hat{j}_i^{\text{spin}(k)}|u_m(\bm{k})\rangle \langle
    u_m(\bm{k})|\Hat{v}_j|u_n(\bm{k})\rangle}{E_{mn}(E_{mn}-\hbar\omega-i\eta)}, 
  \label{eq:spin-dependent-opt-cond}
\end{align}
with
\begin{equation}
  f_{nm}\equiv f(E_n(\bm{k}))-f(E_m(\bm{k})),
\end{equation}
and 
\begin{equation}
  E_{mn}\equiv E_m(\bm{k})-E_n(\bm{k}),
\end{equation}
where 
$n(m)$ indicates the band index including spin degree of freedom and $|u_n(\bm{k}) \rangle$ is the eigen function of the eigen energy $E_n(\bm{k})$. $f(E_n(\bm{k}))$ represents the Fermi-Dirac distribution function which is given by
\begin{equation}
  f(E_n(\bm{k})) = \left({1+\exp\left[\frac{E_n(\bm{k})-\mu}{k_BT}\right]}\right)^{-1},
  \label{eq:fermi-dirac}
\end{equation}
where $k_B$ is the Boltzmann constant and $T$ is the temperature.
$\eta$ is an infinitesimally small real number. Throughout this paper,
$\eta = 0.001$ eV is set for the calculation of conductivity.  
Note that the specific k-point (1000 $\times$ 1000) and the chosen
$\eta$ value were determined by convergence tests. 
$\int_{\text{BZ}}$ denotes integration over the first BZ. 
$\hat{j}_{i}^{\text{spin}(k)}(\bm{k})$ is the $\bm{k}$-dependent
spin current operator, which is defined as 
\begin{equation}
  \hat{j}_{i}^{\text{spin}(k)}(\bm{k})=\frac{1}{2}\left\{\frac{\hbar}{2}\hat{\sigma}_k\otimes\hat{I}_6,\hat{v}_i(\bm{k})\right\},  
\end{equation}
which is given by the anticommutation between the group velocity
operator
$\hat{v}_i=\frac{1}{\hbar}\frac{\partial\hat{H}}{\partial{k_i}}$ and
$k$-component of the Pauli matrix $\hat{\sigma}_k$ ($k=x,y,z$). Here,
$\hat{I}_6$ is the $6\times 6$ identity matrix. 
Equation (\ref{eq:spin-dependent-opt-cond}) can be rewritten as
\begin{align}
  \sigma_{ij}^{\text{spin}(k)}(\omega)=\frac{e}{{(2\pi)}^2}\int_{\text{BZ}}\Omega^{\text{spin}(k)}_{ij}(\omega,\bm{k})d^2\bm{k}
  \label{eq:spin-dependent-opt-cond-2}
\end{align}
with
\begin{align}
  &\Omega^{\text{spin}(k)}_{ij}(\omega,\bm{k})=\hbar\sum_{n}f(E_n(\bm{k}))\nonumber\\
  &\sum_{m(\neq n)}\frac{-2\text{Im}\langle u_n(\bm{k})|\Hat{j}_i^{\text{spin}(k)}|u_m(\bm{k})\rangle \langle u_m(\bm{k})|\Hat{v}_j|u_n(\bm{k})\rangle }{E_{mn}(E_{mn}-\hbar\omega-i\eta)},
  \label{eq:spin-berry-curvature-like}
\end{align}
where $\Omega^{\text{spin}(k)}_{ij}(\omega,\bm{k})$ is an integrand of
spin-dependent optical conductivity~\cite{Qiao2018}. 
\begin{figure}
  \centering
  \includegraphics[width=85mm]{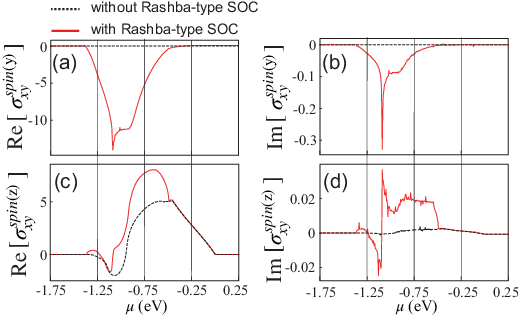} 
  \caption{Fermi energy dependence of spin-dependent Hall conductivities $\sigma_{xy}^{\text{spin}(k)}$, where $k=(y,z)$ indicates the direction of spin polarization of optically generated pure spin current:
    (a) The real part and (b) the imaginary part of $\sigma^{\text{spin}(y)}_{xy}$, respectively.
    (c) The real part and (d) the imaginary part of $\sigma^{\text{spin}(z)}_{xy}$, respectively.
  Here, red (dashed black) lines indicate the result with (without)
  Rashba-type SOC, i.e., WSeTe (WSe$_2$). 
  The unit of $\sigma^{\text{spin}(k)}_{xy}$ is $\frac{e\hbar^2}{4(2\pi)^2}$.
  }
  \label{dc_limit_conductivity}
  \end{figure}

\subsection{DC Limit}
In the limit of direct current (DC), i.e., $\omega=0$ at zero temperature
($T=0K$) of the clean system ($\eta=0$ eV),
Eq. (\ref{eq:spin-berry-curvature-like}) can be rewritten as the spin
Berry curvature~\cite{Habara2021}, i.e.,
\begin{align}
  &\Omega^{\text{spin}(k)}_{i\perp j}(\bm{k})=\hbar\sum_{n}f(E_n(\bm{k}))\nonumber\\
  &\sum_{m(\neq n)}\frac{-2\text{Im}\langle u_n(\bm{k})|\Hat{j}_i^{\text{spin}(k)}|u_m(\bm{k})\rangle \langle u_m(\bm{k})|\Hat{v}_j|u_n(\bm{k})\rangle }{E_{mn}^2}.
  \label{eq:spin-berry-curvature-like-dclimit}
\end{align}
Here, $i\perp j$ indicates that $i$ direction is perpendicular to $j$
direction. 
The spin Berry curvature induces an anomalous velocity 
\begin{align}
  \bm{v}_{\perp} = -\frac{e}{\hbar}\bm{\mathcal{E}}\times\bm{\Omega}^{\text{spin}(k)},
\end{align}
under the presence of an electric field $\bm{\mathcal{E}}$. 
Here, the spin Berry curvature vector is given by $\bm{\Omega}^{\text{spin}(k)}~=~(0,0,\Omega_{i\perp j}^{\text{spin}(k)}(\bm{k}))$.
\begin{figure}
\centering
\includegraphics[width=85mm]{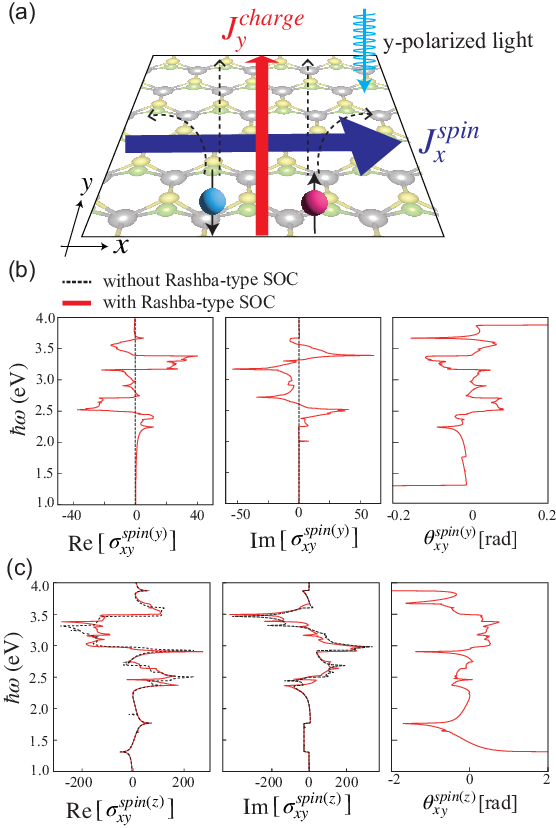} 
\caption{
  (a) Schematic of the generation of spin Hall currents under the
  linearly polarized light irradiation. 
  (b)(c) Angular frequency $\omega$ dependence of spin-dependent 
  optical Hall conductivities $\sigma_{xy}^{\text{spin}(k)}$ and 
  spin Hall angle $\theta_{xy}^{\text{spin}(k)}$ for monolayer WSeTe, 
  where $k=(y,z)$ indicates the direction of spin-polarization of 
  optically generated pure spin current. Here the direction of 
  spin polarization is (b) $y$- and (c) $z$-direction. 
  The left panels of (b) and (c) show the real part of 
  $\sigma_{xy}^{\text{spin}(k)}(\omega)$, while the center panels show
  the imaginary part of $\sigma_{xy}^{\text{spin}(k)}(\omega)$.
  The right panels of (b) and (c) indicate 
  the spin Hall angle $\theta_{xy}^{\text{spin}(k)}$. 
  The red lines show the results with Rashba-type SOC, i.e. WSeTe, 
  while the black dot lines represent the case without it, i.e. WSe$_2$. 
  The unit of $\sigma^{\text{spin}(k)}_{xy}$ is $\frac{e\hbar^2}{4(2\pi)^2}$.
}
\label{spin_xyz_optical_cond}
\end{figure}

Figure~\ref{dc_limit_conductivity} shows the Fermi energy dependence
of spin Hall conductivity for different spin polarized directions, with (red line) and without (black dashed line) Rashba-type SOC.
Since $\sigma_{ij}^{\text{spin}(k)}(\omega)$ is a complex function, we
can separate it into real and imaginary parts as
\begin{equation}
  \sigma_{ij}^{\text{spin}(k)}(\omega) = \text{Re}[\sigma_{ij}^{\text{spin}(k)}(\omega)] + i\text{Im}[\sigma_{ij}^{\text{spin}(k)}(\omega)].
\end{equation}
Figure~\ref{dc_limit_conductivity}(a) shows the real part of spin Hall conductivity, $\text{Re}[\sigma^{\text{spin}(y)}_{xy}]$, which generates the pure spin Hall current with $y$-polarized spin, highlighting that the inclusion of Rashba-type SOC generates in-plane spin-polarized conductivities. 
Figure~\ref{dc_limit_conductivity}(b) displays the corresponding imaginary part, $\text{Im}[\sigma^{\text{spin}(y)}_{xy}]$.
Figure~\ref{dc_limit_conductivity}(c) illustrates the real part of spin Hall conductivity, $\text{Re}[\sigma^{\text{spin}(z)}_{xy}]$, to generate the pure spin Hall current with $z$-polarized spin, which appears even without Rashba-type SOC. Figure~\ref{dc_limit_conductivity}(d) shows the corresponding imaginary part, $\text{Im}[\sigma^{\text{spin}(z)}_{xy}]$.
Although the $z$-polarized pure spin Hall current is generated by the Ising-type SOC, the presence of Rashba-type SOC further enhances it. 

In addition, the other components of the conductivity are calculated
as follows:
$\sigma^{\text{spin}(x)}_{yy}=-\sigma^{\text{spin}(x)}_{xx}$,
$\sigma^{\text{spin}(y)}_{yx}=\sigma^{\text{spin}(y)}_{xy}$,
$\sigma^{\text{spin}(z)}_{yx}= -\sigma^{\text{spin}(z)}_{xy}$ and
$\sigma^{\text{spin}(x)}_{xy}=\sigma^{\text{spin}(x)}_{yx}=\sigma^{\text{spin}(y)}_{xx}=\sigma^{\text{spin}(y)}_{yy}=\sigma^{\text{spin}(z)}_{xx}=\sigma^{\text{spin}(z)}_{yy}=0$.  
In Appendix~\ref{sec_Analysis_of_SBC}, we provide the detail for the
symmetry of Berry curvature and spin Berry curvature in the 1st BZ.

These results reveal the effects of Rashba-type SOC on the generation and
modulation of the spin Hall currents in monolayer Janus TMDCs. 
In TMDCs such as WSe$_2$, where Rashba-type SOC is typically negligible, the spin Hall current exhibits only out-of-plane polarization. In contrast, the broken out-of-plane symmetry in Janus TMDCs such as WSeTe leads to a significant Rashba-type SOC.
Rashba-type SOC plays a crucial role for generating in-plane polarized pure
spin Hall current and enhancement of out-of-plane polarized pure spin
hall current. These results contribute to the development of spintronic
devices. Notably, the generation of the pure spin Hall current is further enhanced by hole doping. 

\subsection{Optical Conductivity}
Next, we examine the angular frequency dependence of
$\sigma_{ij}^{\text{spin}(k)}(\omega)$ for monolayer WSeTe. 
Figure~\ref{spin_xyz_optical_cond}(a) schematically illustrates the numerical
results that the optical generation of pure spin Hall current
$J_x^{\text{spin}}$ in $x$-direction
under the irradiation of $y$-polarized light which drives a charge
current along the $y$-direction.
The left and center panels of Fig.~\ref{spin_xyz_optical_cond}(b) show
the real and imaginary part of spin-dependent optical Hall conductivity $\sigma^{\text{spin}(y)}_{xy}(\omega)$, i.e., the generation of spin Hall current with $y$-polarized spin.
The red lines show the results with Rashba-type SOC, while the black dot lines represent the case without it.
This result indicates that the Rashba-type SOC plays a crucial 
role for generating in-plane spin-polarized current by irradiating
linearly polarized light. 
The left and center panels of Fig.~\ref{spin_xyz_optical_cond}(c) illustrate the real and imaginary part of spin-dependent optical
Hall conductivity $\sigma^{\text{spin}(z)}_{xy}(\omega)$, 
i.e., the generation of spin Hall current with $z$-polarized spin.
In this case, the conductivity remains finite even without Rashba-type
SOC due to the presence of Ising-type SOC. 

By considering crystal symmetry and Neumann's principle, we can obtain
the other components of conductivity tensor as follows:
$\sigma^{\text{spin}(x)}_{yy}=-\sigma^{\text{spin}(x)}_{xx}$,
$\sigma^{\text{spin}(y)}_{yx}=\sigma^{\text{spin}(y)}_{xy}$,
$\sigma^{\text{spin}(z)}_{yx}= -\sigma^{\text{spin}(z)}_{xy}$ and
$\sigma^{\text{spin}(x)}_{xy}=\sigma^{\text{spin}(x)}_{yx}=\sigma^{\text{spin}(y)}_{xx}=\sigma^{\text{spin}(y)}_{yy}=\sigma^{\text{spin}(z)}_{xx}=\sigma^{\text{spin}(z)}_{yy}=0$.
The detail is summarized in Appendix~\ref{sec_Neumann}.

Here we shall pay attention to the polarization direction of the generated
spin Hall current.
In the conventional SHE, a transverse spin current
$J_i^{\text{spin}(k)}$ is induced in response to a charge current
$J_j^{\text{charge}}$, where the spin polarization direction $k$ is
perpendicular to both, i.e., $i\perp j\perp k$ ($i,j=x,y,z$).
However, the spin Hall currents in WSeTe are not constrained by this
conventional relation. 
We classify these SHEs into the following types:
(i) conventional SHE, where $J_i^{\text{spin}(k)}$,
$J_j^{\text{charge}}$ and $k$ are mutually orthogonal, i.e., $i\perp j
\perp k$, corresponding to Fig.~\ref{spin_xyz_optical_cond}(c);
(ii) in-plane SHE (IPSHE), where $J_i^{\text{spin}(k)}$ and
$J_j^{\text{charge}}$ are transverse, but $k$ is parallel to one of
them, i.e., $i \perp j \parallel k$ or $i \parallel k \perp j$,
corresponding to Fig.~\ref{spin_xyz_optical_cond}(b). 

Furthermore, we consider the angular frequency dependence of the spin
Hall angle (SHA)~\cite{Tao2018}, which measures the conversion
efficiency from charge current to spin current. 
SHA can be given as
\begin{equation}
    \theta_{ij}^{\text{spin}(k)} = \frac{2e}{\hbar} \frac{{\rm
        Re}[\sigma_{ij}^{\text{spin}(k)}]}{{\rm Re}[\sigma^{\text{charge}}_{jj}]},
\end{equation}
where $\sigma^{\text{spin}(k)}_{ij}$ represents the spin-dependent
optical conductivity, $\sigma^{\text{charge}}_{jj}$ is the optical
longitudinal conductivity. 
The factor $\frac{2e}{\hbar}$ ensures that the units of spin conductivity match those of charge conductivity.
The right panels of Figs.~\ref{spin_xyz_optical_cond}(b) and (c) show SHA of
monolayer WSeTe for different spin polarization direction: (b)
$y$-direction $\theta_{xy}^{\text{spin}(y)}$ and (c) $z$-direction
$\theta_{xy}^{\text{spin}(z)}$.  
Each SHA has divergence at approximately 1.2 eV and 3.9 eV due to
$\sigma^{\text{charge}}_{yy}$ becoming zero, making SHA
ill-defined. When the spin-dependent optical conductivity reaches a 
peak in the left panels of Figs.~\ref{spin_xyz_optical_cond}(b) and (c), the
SHA also increases, indicating that spin current can be generated
efficiently. 

In Appendix~\ref{sec_circular-polarized-light}, 
we derive the optically generated spin current by circularly
polarized light irradiation. 
%
In Appendix~\ref{sec_temperature_dependence}, 
we examine the temperature dependence of the spin-dependent 
optical Hall conductivity. 
Our finding indicates that the spin-dependent optical Hall conductivity is 
robust against temperature variations and is expected to be observable even 
at room temperature.

\section{conclusion}\label{sec_conclusion}
In this study, we have theoretically demonstrated the generation of
pure spin Hall current in monolayer WSeTe under linearly polarized
light irradiation.
Monolayer WSeTe possesses both Ising-type SOC and Rashba-type SOC due
to broken out-of-plane mirror symmetry. 
These SOCs induce the finite spin Berry curvature, resulting in the
generation of the spin Hall effect by light irradiation.
In particular, the presence of Rashba-type SOC induces in-plane
polarized spin-dependent optical conductivities, giving rise to
unconventional spin Hall effects,. i.e., IPSHE. 
IPSHE enables direct control of the spin polarization direction within
the plane of the monolayer. 
Thus, monolayer WSeTe can be utilized for the source of pure spin Hall
current and next-generation opt-spintronics devices such as in-plane
spin injection, detection and manipulation under linearly polarized
light using IPSHE, where conventional SHE is less optimal. 
In addition, we have calculated SHA, which measures conversion
efficiency from charge current to spin Hall current. The results show
that each SHA is enhanced by irradiating visible light.  

Our results offer a new degree of freedom for designing optospintronic
devices, such as spin current harvesting via light irradiation in 2D
materials. 

\section*{ACKNOWLEDGMENTS}
This work was supported by JSPS KAKENHI (Grants No. JP25K01609, No. JP22H05473, and No. JP21H01019), JST CREST (Grant No. JPMJCR19T1). K. W. acknowledges the financial support for Basic Science Research Projects (Grant No. 2401203) from the Sumitomo Foundation. Sasakawa Scientific Research Grant from The Japan Science Society.

\section*{Data Availability}
The data that support the findings of this study are available from the corresponding author upon reasonable request \cite{kameda2025data}.

\appendix
\section{Matrix Elements of Monolayer WSeTe}\label{sec_appendix_TBM}
In Sec.~\ref{sec_Electric properties}, we introduced the effective
hamiltonian of monolayer WSeTe given by Eq.(\ref{eq:H_TNN}).
Here we provide the details of
matrix elements, i.e., $V_0, V_1, V_2, V_{11}, V_{12},$ and
$V_{22}$. These elements are expressed as follows.
\begin{widetext} 
\begin{align}
  V_{0} &= \epsilon_1 + 2t_0(2\cos\alpha \cos\beta + \cos 2\alpha)
  + 2r_0(2\cos 3\alpha \cos\beta + \cos 2\beta)
  + 2u_0(2\cos 2\alpha \cos 2\beta + \cos 4\alpha),\nonumber \\
  \text{Re}[V_1] &= -2\sqrt{3}t_2 \sin\alpha \sin\beta
  + 2(r_1 + r_2) \sin 3\alpha \sin\beta
  - 2\sqrt{3}u_2 \sin 2\alpha \sin 2\beta,\nonumber \\
  \text{Im}[V_1] &= 2t_1 \sin\alpha(2\cos\alpha + \cos\beta)
  + 2(r_1 - r_2) \sin 3\alpha \cos\beta
  + 2u_1 \sin 2\alpha (2\cos 2\alpha + \cos 2\beta),\nonumber \\
  \text{Re}[V_2] &= 2t_2(\cos 2\alpha - \cos\alpha \cos\beta)
  - \frac{2}{\sqrt{3}} (r_1 + r_2)(\cos 3\alpha \cos\beta - \cos
  2\beta) + 2u_2(\cos 4\alpha - \cos 2\alpha \cos 2\beta),\nonumber \\ 
  \text{Im}[V_2] &= 2\sqrt{3}t_1 \cos\alpha \sin\beta
  + \frac{2}{\sqrt{3}} (r_1 - r_2) \sin\beta (\cos 3\alpha + 2\cos\beta)
  + 2\sqrt{3}u_1 \cos 2\alpha \sin 2\beta,\nonumber \\
  V_{11} & = \epsilon_2 + (t_{11} + 3t_{22}) \cos\alpha \cos\beta
  + 2t_{11} \cos 2\alpha + 4r_{11} \cos 3\alpha \cos\beta \nonumber \\
 & + 2(r_{11} + \sqrt{3}r_{12}) \cos 2\beta
  + (u_{11} + 3u_{22}) \cos 2\alpha \cos 2\beta
  + 2u_{11} \cos 4\alpha,\nonumber \\
  \text{Re}[V_{12}] &= \sqrt{3}(t_{22} - t_{11}) \sin\alpha \sin\beta
  + 4r_{12} \sin 3\alpha \sin\beta
  + \sqrt{3}(u_{22} - u_{11}) \sin 2\alpha \sin 2\beta,\nonumber \\
  \text{Im}[V_{12}] &= 4t_{12} \sin\alpha(\cos\alpha - \cos\beta)
  + 4u_{12} \sin 2\alpha(\cos 2\alpha - \cos 2\beta),\nonumber\\
  V_{22} &= \epsilon_2 + (3t_{11} + t_{22}) \cos\alpha \cos\beta
  + 2t_{22} \cos 2\alpha
  + 2r_{11}(2\cos 3\alpha \cos\beta + \cos 2\beta)& \nonumber\\
 & + \frac{2}{\sqrt{3}}r_{12}(4\cos 3\alpha \cos\beta - \cos 2\beta)
  + (3u_{11} + u_{22}) \cos 2\alpha \cos 2\beta + 2u_{22} \cos 4\alpha.\nonumber
\end{align}
\end{widetext}
The specific hopping parameters in this TBM can be given as:
\begin{equation}
    E^{jj'}_{\mu\mu'}(\mathbf{R}) = \langle \phi^j_\mu(\mathbf{r}) | \hat{H}(\bm{k}) | \phi^{j'}_{\mu'}(\mathbf{r} - \mathbf{R}) \rangle,
\end{equation}
where $|\phi^j_\mu\rangle$ indicates an atomic orbital of the W atom, and in this paper, we consider 
$|\phi^1_1\rangle~=~d_{z^2}$, $|\phi^2_1\rangle~=~d_{xy}$, and $|\phi^2_2\rangle~=~d_{x^2-y^2}$. 
Here, we define $t_0~=~E^{11}_{11}(\mathbf{R}_1)$, $t_1~=~E^{12}_{11}(\mathbf{R}_1)$, 
$r_0~=~E^{11}_{11}(\tilde{\mathbf{R}}_1)$, $r_1~=~E^{12}_{11}(\tilde{\mathbf{R}}_1)$, 
$u_0~=~E^{11}_{11}(2\mathbf{R}_1)$, and $u_1~=~E^{12}_{11}(2\mathbf{R}_1)$. 
The fitted parameters are summarized in Table~\ref{appendix_table_parameter}.
\begin{table*}
  \begin{center}
  \caption{Parameters for effective TBM of monolayer WSeTe. The unit
    of all the parameters is eV.}
   \begin{tabular}{@{}cccccccccccc@{}} \hline \hline
  &$\epsilon_1$ & $\epsilon_2$ & $t_0$ & $t_1$ & $t_2$ & $t_{11}$ & $t_{12}$ & $t_{22}$ & $r_0$ & $r_1$ & $r_{2}$ \\
  &$r_{11}$ &$r_{12}$ &$u_0$ &$u_1$ &$u_2$ &$u_{11}$ &$u_{12}$ &$u_{22}$ & $\lambda_{\text{SOC}}$ &$\alpha_{0}$ \\ \hline
  WSeTe& 0.728 & 1.655 & -0.146 & -0.124 & 0.507 & 0.117 & 0.127 & 0.015 & 0.036 & -0.234 &0.107 \\
  & 0.044& 0.075& -0.061& 0.032& 0.007& 0.329& -0.202& -0.164& 0.228 & 0.045 \\ \hline
  \end{tabular}
  \label{appendix_table_parameter}
  \end{center}
  \end{table*}

\section{Symmetry Analysis of the Integrands for charge and spin-dependent Conductivities}\label{sec_Analysis_of_SBC}
\begin{figure}
  \centering
  \includegraphics[width=75mm]{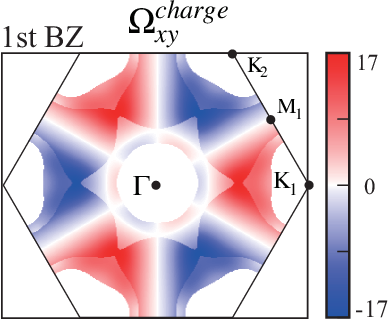}
  \caption{Contour plot of Berry curvature
    $\Omega^{\text{charge}}_{xy}$ for WSeTe in first BZ at $E=-0.85$
    eV.  
  }\label{appendix_bc_dclimit}
\end{figure}

In Sec.~\ref{sec_spin-dependent-cond}, we calculated spin-dependent
optical conductivities, integrating
Eq.~(\ref{eq:spin-berry-curvature-like}) over the first BZ. In a
similar manner, here we shall focus on the integrand of the charge
conductivity. 
In the DC limit, the integrand can be represented as 
\begin{align}
  &\Omega^{\text{charge}}_{ij}(\bm{k})=\hbar\sum_{n}f(E_n(\bm{k}))\nonumber\\
  &\sum_{m(\neq n)}\frac{-2\text{Im}\langle u_n(\bm{k})|\Hat{v}_i|u_m(\bm{k})\rangle \langle u_m(\bm{k})|\Hat{v}_j|u_n(\bm{k})\rangle }{E_{mn}^2}.
  \label{eq:berry-curvature}
\end{align}
For $i\perp j$, this is nothing more than the Berry curvature. 
Figure~\ref{appendix_bc_dclimit} provides counter plots of $\Omega^{\text{charge}}_{xy}$, i.e., the Berry curvature, at an energy level of -0.85 eV.
The point-group symmetry of the berry curvature is $C_{6v}$ and the Mulliken notations for irreducible representation (IR) is $B_1$.
Notably, $B_{1}$ indicates that the berry curvature antisymmetric under
the symmetry operations of the point group, vanishing of the total
integral over the BZ.
Thus, the charge Hall conductivity becomes identically zero, although
this is trivial from that the system has the time-reversal symmetry.

\begin{figure*}
  \centering
  \includegraphics[width=170mm]{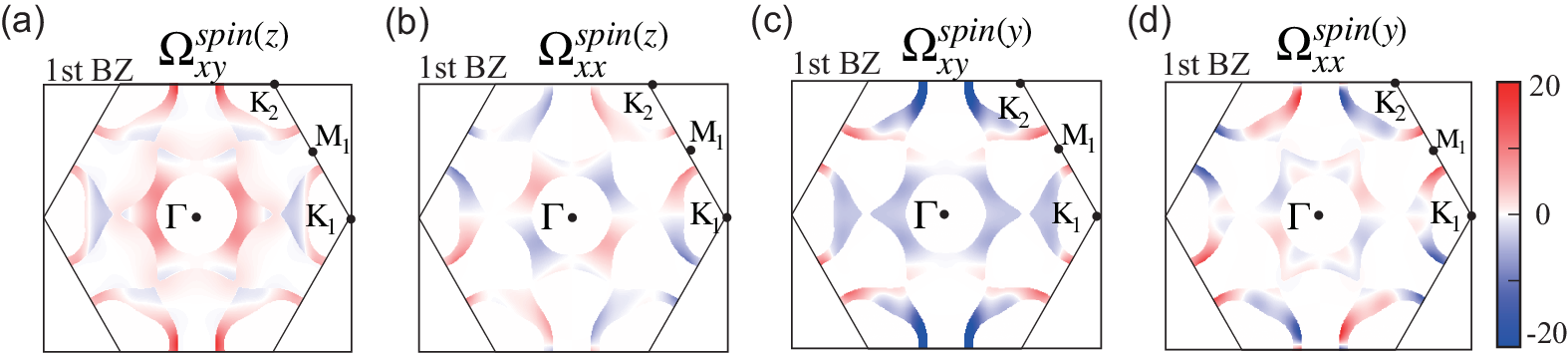}
  \caption{Contour plot of spin Berry curvature for WSeTe in the first
    BZ at $E=-0.85$ eV.
    Spin Berry curvature of $z$-direction
    spin-polarization: (a) $\Omega_{xy}^{\text{spin(z)}}$ and (b)
    $\Omega_{xx}^{\text{spin(z)}}$. Spin Berry curvature of $y$-direction spin-polarization: 
    (c) $\Omega_{xy}^{\text{spin(y)}}$ and (d) $\Omega_{xx}^{\text{spin(y)}}$.
  }\label{appendix_sbc_dclimit}
\end{figure*}
Similarly, we can analyze the symmetry of spin Berry curvature in the
first BZ.
Figure~\ref{appendix_sbc_dclimit}(a) shows
$\Omega_{xy}^{\text{spin(z)}}$, i.e., the spin Berry curvature for
spin polarized along the $z$-axis. 
Figure~\ref{appendix_sbc_dclimit}(b) shows
$\Omega_{xx}^{\text{spin(z)}}$, which is the integrand of longitudinal
spin-dependent conductivity for spin polarized along the $z$-axis.  
Figure~\ref{appendix_sbc_dclimit}(c) shows
$\Omega_{xy}^{\text{spin(y)}}$, i.e., the spin Berry curvature for
spin polarized along the $y$-axis.  
Figure~\ref{appendix_sbc_dclimit}(d) shows
$\Omega_{xx}^{\text{spin(y)}}$, which is the integrand of longitudinal
spin-dependent conductivity for spin polarized along the $y$-axis.  
The charge and spin-dependent conductivities are obtained by
integrating these integrands over the first BZ.

Table~\ref{supple_table_symmetry} shows the point-group symmetries of
the spin Berry curvatures presented in
Fig.~\ref{appendix_sbc_dclimit}. In this table, $A_{1}$ and $A_{2}$
represent the Mulliken notations for IRs of the relevant point
groups. Notably, $A_{2}$ indicates that spin Berry curvatures becomes
antisymmetric under the symmetry operations, leading to 
the vanishing the total integral over the BZ.  
\begin{table*}
  \begin{center}
  \caption{Point-group symmetries of the spin Berry curvatures
    corresponding to the irreducible representations (IRs); $A_1$ and
    $A_2$. The IR $A_2$ indicates that spin Berry curvatures becomes the antisymmetric under the symmetry
    operations of the relevant point-group. Thus, the total
    integral of these antisymmetric components over the BZ vanishes.} 
   \begin{tabular}{@{}ccccc@{}} \hline \hline
  & (a) $\Omega^{\text{spin(z)}}_{xy}$ & (b)$\Omega^{\text{spin(z)}}_{xx}$ 
  & (c)$\Omega^{\text{spin}(y)}_{xy}$ & (d)$\Omega^{\text{spin}(y)}_{xx}$\\\hline
  point group&$C_{2v}$ &$C_{2v}$ &$C_{2v}$ &$C_{2v}$ \\
  IRs&$A_1$ &$A_2$ &$A_1$ &$A_2$ \\ \hline
  Integrated value over BZ&Finite &0 & Finite &0 \\ \hline
  \end{tabular}
  \label{supple_table_symmetry}
  \end{center}
  \end{table*}

\begin{figure}
  \centering
  \includegraphics[width=75mm]{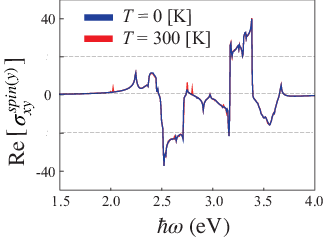}
  \caption{
    Temperature dependence of spin-dependent optical Hall
    conductivity $\text{Re}[\sigma^{\text{spin}(y)}_{xy}(\omega)]$ for monolayer WSeTe.
    The blue line shows the result at $T=0$ K, while the red line
    represents the result at $T=300$ K.
  }\label{appendix_temperature_dependence} 
\end{figure}

\section{Neumann's Principle Analysis on Spin-Dependent Conductivity Tensor}\label{sec_Neumann}
In this section, we discuss the elements of spin-dependent
conductivity tensor by using the Neumann's principle.
According to Neumann's principle, the symmetry group governing any
physical property of a crystal must include all symmetry operations of
the crystal's point group. 
Note that because the spin operator is a pseudovector while the
velocity operator is a polar vector, their combination yields a
third-rank pseudotensor for the spin-dependent conductivity, i.e.,
spin-dependent conductivity tensor
$\sigma^{\text{spin}(\alpha)}_{\beta \gamma}$ can be represented a
third rank pseudo-tensor ($\alpha, \beta, \gamma = x, y, z$). 

The crystal symmetry of monolayer TMDCs with the point group $D_{3h}$
is characterized by a three-fold rotational axis ($C_3$) perpendicular
to the plane of the layers and a horizontal mirror plane
($M_{h}$) that lies parallel to the layers. 
In contrast, monolayer Janus TMDCs break the out-of-plane mirror symmetry 
due to the different calcogens on each side of the transition metal layer, 
changing the point group to $C_{3v}$. 
The point group $C_{3v}$ retains the three-fold rotational symmetry 
but lacks the horizontal mirror plane symmetry.
The full set of symmetry operations for $C_{3v}$ is:\{$\hat{E},
\hat{C_3}, \hat{C_{3}^2}, \hat{M_{v}}, \hat{M_{v}^\prime},
\hat{M_{v}^{\prime\prime}}$\}, where $\hat{E}$ is the identity
operation, $\hat{C_3}$ is the rotation by $\frac{2\pi}{3}$ around the
$z$ axis, $\hat{M_{v}}$ is the reflection by the plane perpendicular to the $xy$ plane and through the angular bisector of $R_1$ and $R_4$ in Fig.~\ref{janus_tmdc_structure}(a), and $\hat{M_{v}^\prime}$ and $\hat{M_{v}^{\prime\prime}}$ are obtained by rotating $\hat{M_{v}}$ around $z$ axis by $\frac{2\pi}{3}$ and $\frac{4\pi}{3}$ respectively.

The spin-dependent conductivity tensor can be represented as 
\begin{equation}
  \bm{\sigma}^{\text{spin}(\alpha)}=
  \begin{pmatrix}
    \sigma^{\text{spin}(\alpha)}_{xx} & \sigma^{\text{spin}(\alpha)}_{xy} & \sigma^{\text{spin}(\alpha)}_{xz}\\
    \sigma^{\text{spin}(\alpha)}_{yx} & \sigma^{\text{spin}(\alpha)}_{yy} & \sigma^{\text{spin}(\alpha)}_{yz}\\
    \sigma^{\text{spin}(\alpha)}_{zx} & \sigma^{\text{spin}(\alpha)}_{zy} & \sigma^{\text{spin}(\alpha)}_{zz}
    \end{pmatrix},
  \nonumber
\end{equation}
where $\alpha(=x,y,z)$ indicates the direction of spin polarization.
According to the Neumann's principle, the spin-dependent conductivity
tensor can be expressed in terms of the symmetry operation matrix as $R$, 
\begin{align}
\sigma^{\text{spin}(\alpha)}_{\beta\gamma} =
\text{det}({R})R_{\alpha\alpha^{\prime}}R_{\beta\beta^{\prime}}R_{\gamma\gamma^{\prime}}\sigma^{\text{spin}(\alpha^{\prime})}_{\beta^{\prime}\gamma^{\prime}},
\nonumber
\end{align}
det($R$) is a manifestation of the psudotensor. 
By considering the symmetry of the crystal, we can reduce the number of independent elements in the
spin-dependent conductivity tensor.
Especially, Janus TMDCs have the following symmetries: (1)
$\hat{M_{v}}$(a vertical mirror plane with the mirror axis along the
$x$-axis) and (2) $\hat{C_3}$ (a threefold rotational symmetry).
Since the operator of $\hat{M_{v}}$ is written as
\begin{align}
\hat{M}_v = {\rm diag}(-1,1,1),\nonumber
\end{align}
the spin-dependent conductivity tensor can be simplified as
    \begin{align}
    \bm{\sigma}^{\text{spin}(x)}=
    \begin{pmatrix}
      \sigma^{\text{spin}(x)}_{xx} & 0 & 0 \\
       0 & \sigma^{\text{spin}(x)}_{yy} & 0 \\
       0 & 0 & \sigma^{\text{spin}(x)}_{zz}
    \end{pmatrix},
    \nonumber
    \end{align}
    \begin{align}
    \bm{\sigma}^{\text{spin}(y)}=
      \begin{pmatrix}
        0 & \sigma^{\text{spin}(y)}_{xy} & \sigma^{\text{spin}(y)}_{xz} \\
        \sigma^{\text{spin}(y)}_{yx} & 0 & 0 \\
        \sigma^{\text{spin}(y)}_{zx} & 0 & 0
      \end{pmatrix},
      \nonumber
    \end{align}
    \begin{align}
    \bm{\sigma}^{\text{spin}(z)}=
      \begin{pmatrix}
        0 & \sigma^{\text{spin}(z)}_{xy} & \sigma^{\text{spin}(z)}_{xz} \\
        \sigma^{\text{spin}(z)}_{yx} & 0 & 0 \\
        \sigma^{\text{spin}(z)}_{zx} & 0 & 0
      \end{pmatrix}.
      \nonumber
    \end{align}

Similarly, since
     operator of $\hat{C_3}$ is written as
\begin{align}
     R=\hat{C_3}=
  \begin{pmatrix}
    -\frac{1}{2} & -\frac{\sqrt{3}}{2} & 0 \\
    \frac{\sqrt{3}}{2} & -\frac{1}{2} & 0 \\
     0 & 0 & 1
    \end{pmatrix}, \nonumber
\end{align}
we can obtain the following relations:
\begin{align}
    \begin{cases}
      3\sigma^{\text{spin}(x)}_{xx}+\sigma^{\text{spin}(x)}_{yy}+\sigma^{\text{spin}(y)}_{xy}+\sigma^{\text{spin}(y)}_{yx}=0, \\
      \sigma^{\text{spin}(x)}_{xx}-\sigma^{\text{spin}(x)}_{yy}-\sigma^{\text{spin}(y)}_{xy}+3\sigma^{\text{spin}(y)}_{yx}=0, \\
      \sigma^{\text{spin}(x)}_{xx}+3\sigma^{\text{spin}(x)}_{yy}-\sigma^{\text{spin}(y)}_{xy}-\sigma^{\text{spin}(y)}_{yx}=0.
   \end{cases} \nonumber
\end{align}
Thus, the following relation is deduced,
\begin{align}
   \sigma^{\text{spin}(x)}_{xx}=-\sigma^{\text{spin}(x)}_{yy}=-\sigma^{\text{spin}(y)}_{xy}=-\sigma^{\text{spin}(y)}_{yx}.
\nonumber
\end{align}
Consequently, the spin-dependent conductivity tensor can be represented as
\begin{align}
  \bm{\sigma}^{\text{spin}(x)}=
  \begin{pmatrix}
    \sigma^{\text{spin}(x)}_{xx} & 0\\
    0 & -\sigma^{\text{spin}(x)}_{xx}
  \end{pmatrix},
  \label{eq:spin-conductivity-tensor_x}
\end{align}
  \begin{align}
  \bm{\sigma}^{\text{spin}(y)}=
  \begin{pmatrix}
    0 & -\sigma^{\text{spin}(x)}_{xx} \\
    -\sigma^{\text{spin}(x)}_{xx} & 0
  \end{pmatrix},
  \label{eq:spin-conductivity-tensor_y}
\end{align}
  \begin{align}
  \bm{\sigma}^{\text{spin}(z)}=
  \begin{pmatrix}
    0 & \sigma^{\text{spin}(z)}_{xy}\\
    \sigma^{\text{spin}(z)}_{yx} & 0
  \end{pmatrix},
  \label{eq:spin-conductivity-tensor_z}
\end{align}
where the conductivity components in the $z$-direction are omitted,
because we are considering a 2D system. 
Here, the conditions for these tensor elements are consistent with our
numerical calculations of spin-dependent conductivities.

\section{Spin Current Induced by Circularly Polarized Light}
\label{sec_circular-polarized-light}
Here we extend our symmetry analysis to the case of circularly
polarized light~\cite{Freimuth2021, Akita2020, Xu2021}.  
Spin current in the $i$ direction with spin component $k$ can be given by~\cite{Xu2021}
\begin{align}
  {J}^{\text{spin}(k)}_i(\omega) = \int_{\text{BZ}}\frac{d^2\bm{k}}{(2\pi)^2}\sum_{nm}\frac{f_{nm} J_{i,nm}^{\text{spin}(k)} V_{mn}(\omega)}{E_{mn}-\hbar\omega-i\eta},
  \nonumber
\end{align}
where $J_{i,nm}^{\text{spin}(k)}\equiv \langle u_n(\bm{k})|\hat{j}_i^{\text{spin}(k)}|u_m(\bm{k})\rangle$.
The interaction matrix element, $V_{mn}(\omega)$, is given by
\begin{align}
  V_{mn}(\omega) = \frac{ie}{\omega}\langle u_m(\bm{k})|\hat{\bm{v}}\cdot \bm{\mathcal{E}}|u_n(\bm{k})\rangle,
  \nonumber
\end{align}
where $\hat{\bm{v}}=(\hat{v}_x, \hat{v}_y)$ is the group velocity operator vector and $\bm{\mathcal{E}}=(\mathcal{E}_x, \mathcal{E}_y)$ is an external electric field.
Therefore, the spin current can be rewritten as
\begin{align}
  {J}^{\text{spin}(k)}_i(\omega)&=\frac{i\hbar e}{{(2\pi)}^2}\int_{\text{BZ}}d^2\bm{k}\sum_{n\neq m}f_{nm} \nonumber\\
  &\frac{J^{\text{spin}(k)}_{i,nm}\langle u_m(\bm{k})|\Hat{v}_x\mathcal{E}_x+\Hat{v}_y\mathcal{E}_y|u_n(\bm{k})\rangle}{E_{mn}(E_{mn}-\hbar\omega-i\eta)} \nonumber\\
  &=\sigma^{\text{spin}(k)}_{ix}(\omega)\mathcal{E}_x+\sigma^{\text{spin}(k)}_{iy}(\omega)\mathcal{E}_y.
  \nonumber
\end{align}
For a right-handed circularly polarized (RCP) light,
the electric field can be expressed as
\begin{equation}
  \bm{\mathcal{E}}_{+} = \frac{1}{\sqrt{2}} \mathcal{E}_0(1, -i).
  \nonumber
\end{equation}
Similarly, for left-handed circularly polarized (LCP) light, the
electric field is expressed as
\begin{equation}
  \bm{\mathcal{E}}_{-} = \frac{1}{\sqrt{2}}\mathcal{E}_0(1,i).
  \nonumber
\end{equation}
Then, the spin current flowing along $i=x, y$ direction with the spin polarization $k$-axis can be expressed as
\begin{equation}
  J^{\text{spin}(k)}_{x,\pm}(\omega) = \frac{\mathcal{E}_0}{\sqrt{2}}\text{Re}[\sigma^{\text{spin}(k)}_{xx}(\omega)\mp i\sigma^{\text{spin}(k)}_{xy}(\omega)], 
  \nonumber
\end{equation}
\begin{equation}
  J^{\text{spin}(k)}_{y,\pm}(\omega) = \frac{\mathcal{E}_0}{\sqrt{2}}\text{Re}[\sigma^{\text{spin}(k)}_{yx}(\omega)\mp i\sigma^{\text{spin}(k)}_{yy}(\omega)], 
  \nonumber
\end{equation}
where the subscript $+$ ($-$) sign corresponds to RCP (LCP) and $\mathcal{E}_0$ is the amplitude of the electric field.

In Janus TMDCs, the spin-dependent conductivity tensor satisfies the relations given in Eqs.~{(\ref{eq:spin-conductivity-tensor_x})-(\ref{eq:spin-conductivity-tensor_z})}. 
Consequently, the spin current can be rewritten in terms of the spin-dependent conductivity tensor as follows:
\begin{align}
  J^{\text{spin}(x)}_{x,\pm}(\omega) &= \frac{\mathcal{E}_0}{\sqrt{2}}\text{Re}[\sigma^{\text{spin}(x)}_{xx}(\omega)],\nonumber\\
  J^{\text{spin}(x)}_{y,\pm}(\omega) &= \pm\frac{\mathcal{E}_0}{\sqrt{2}}\text{Re}[i\sigma^{\text{spin}(x)}_{xx}(\omega)],\nonumber\\
  J^{\text{spin}(y)}_{x,\pm}(\omega) &= \pm\frac{\mathcal{E}_0}{\sqrt{2}}\text{Re}[i\sigma^{\text{spin}(x)}_{xx}(\omega)],\nonumber\\
  J^{\text{spin}(y)}_{y,\pm}(\omega) &= -\frac{\mathcal{E}_0}{\sqrt{2}}\text{Re}[\sigma^{\text{spin}(x)}_{xx}(\omega)],\nonumber\\
  J^{\text{spin}(z)}_{x,\pm}(\omega) &= \mp\frac{\mathcal{E}_0}{\sqrt{2}}\text{Re}[i\sigma^{\text{spin}(z)}_{xy}(\omega)],\nonumber\\
  J^{\text{spin}(z)}_{y,\pm}(\omega) &= \frac{\mathcal{E}_0}{\sqrt{2}}\text{Re}[\sigma^{\text{spin}(z)}_{yx}(\omega)].\nonumber\\
  \nonumber
\end{align}
This approach provides a systematic method to predict the behavior of the spin current for various polarization states.

\section{Temperature Dependence of Spin-Dependent Optical Hall Conductivity}
\label{sec_temperature_dependence}
In this section, we discuss the temperature dependence of the
spin-dependent optical Hall conductivity.
Figure~\ref{appendix_temperature_dependence} shows the temperature
dependence of the real part of spin-dependent optical Hall
conductivity, $\text{Re}[\sigma^{\text{spin}(y)}_{xy}(\omega)]$, for
monolayer WSeTe at two different temperatures. The blue line
represents the result at $T=0$~K, while the red line corresponds to
$T=300$~K.
Our results demonstrate that the spin-dependent optical Hall
conductivity is robust against temperature, indicating that
it can be observable even at room temperature. 

\bibliography{reference}
\end{document}